# Asymmetry of the spectral lines of the coronal hole and quiet Sun in the transition region

Razieh Hosseini ,[1]⋆ Pradeep Kayshap ,[2] Nasibe Alipour  and Hossein Safari [1]⋆

[1]*Department of Physics, Faculty of Science, University of Zanjan, 38791-45371 Zanjan, Iran*
[2]*School of Advanced Sciences and Languages, VIT Bhopal University, Kothrikalan, Sehore, Madhya Pradesh 466114, India*
[3]*Department of Physics, University of Guilan, Rasht 41335-1914, Iran*

## ABSTRACT

The asymmetry of line profiles, i.e. the secondary component, is crucial to understanding the energy release of coronal holes (CH), quiet Sun (QS), and bright points (BPs). We investigate the asymmetry of Si IV 1393.75 Å of the transition-region (TR) line recorded by Interface Region Imaging Spectrograph (IRIS) and co-spatial-temporal Atmospheric Imaging Assembly (AIA) and Helioseismic and Magnetic Imager (HMI) data onboard Solar Dynamics Observatory (SDO) for three time series on 2015 April 26, 2014 July 24, and 2014 July 26. Most asymmetric profiles are in the complex magnetic field regions of the networks. The asymmetric profiles are fitted with both single and double Gaussian models. The mean value of Doppler velocity of the second component is almost zero (with a significant standard deviation) in QS/CH, which may indicate that the physical process to trigger the secondary Gaussian originates at the formation height of Si IV. While the mean Doppler velocity from secondary Gaussian in BPs is around $+4.0\,\mathrm{km\,s^{-1}}$ (redshifted). The non-thermal velocities of the secondary Gaussian in all three regions are slightly higher than the single Gaussian. The statistical investigation leads to the prevalence of blueshifted secondary components in QS/CH. However, secondary Gaussian in the BPs redshifted, i.e. the BPs redshift behaviour could be interpreted due to the site of reconnection located above the formation height of the Si IV line. The peak intensity of the second component for all three regions is likely to follow a power law that is a signature of the small-scale flaring-like trigger mechanism.

## 1 INTRODUCTION

The solar atmosphere contains different types of transients (e.g. jets, filament, flares, coronal mass ejection, etc.) and regions [e.g. active regions (ARs), quiet Sun (QS), coronal holes (CHs), bright points (BPs)]. Due to the low density and temperature, the CHs appear as dark regions in the solar atmosphere. The CHs temporarily occur from the equatorial to the polar regions of the solar atmosphere.

At coronal temperatures, the QS has a more significant emission than CHs, i.e. QS appears bright compared to the CHs. Further, ARs have very stronger emission than QS (Waldmeier 1975; Cranmer 2009; Kayshap, Banerjee & Srivastava 2015; Tripathi, Nived & Solanki 2021). On the contrary, the QS and CH regions appear similar at the chromosphere and photosphere temperatures (see e.g. Stucki et al. 1999, 2000). Kayshap et al. (2018b) studied the intensity differences of CH and QS in the Mg II spectral lines. They showed that the CHs's intensity is lower than QS for regions with similar unsigned magnetic flux. Tripathi, Nived & Solanki (2021) obtained similar findings for QS and CH using C II line emissions. These studies (i.e. Kayshap et al. 2018b; Tripathi, Nived & Solanki 2021) have not reported any visual difference in QS and CH in the solar chromosphere.

A spectroscopic diagnostic is an important tool for investigating *in-situ* plasma of the solar atmosphere, and remotely, we can get valuable information about the plasma (see e.g. Del Zanna & Mason 2018). Depending on the *in-situ* physical conditions, the spectral lines may form in optically thin or thick conditions, which form the shape of spectral profiles. Most lines of the TR and corona are single peak profiles forming in optically thin conditions. Such spectral lines are well characterized by the single Gaussian fit to derive various parameters (e.g. Kayshap, Banerjee & Srivastava 2015; Del Zanna & Mason 2018). However, now high-resolution observations continuously show that these profiles significantly deviate from perfect (single) Gaussian, known as asymmetric profiles. We must say that asymmetric profiles are ubiquitous in TR and corona (e.g. Peter 2000, 2001, 2010; Kayshap et al. 2018a, 2021; Kayshap & Young 2023). The asymmetric spectral profiles may have a narrow core and a broad minor component (about 10 per cent–20 per cent of total emission). The asymmetries of coronal and TR lines have been investigated in several works (e.g. Kjeldseth Moe & Nicolas 1977; Wilhelm et al. 1995; Culhane et al. 2000; Peter 2001, 2010; Hara et al. 2008; McIntosh & De Pontieu 2009a, b). De Pontieu et al. (2009) obtained 5 per cent–10 per cent blueward asymmetry in line profiles of QS and CHs in an important work. Persistent blueward line asymmetries were detected for lines forming at temperatures of $7 \times 10^4$ K to $2 \times 10^6$ K and velocities ranging from 50 to 150 km s$^{-1}$ (De Pontieu et al. 2009; McIntosh & De Pontieu 2009a; Peter 2010). Spectral profiles with asymmetries reveal many details on the characteristics of the generating plasma along the line of sight. The superposition of profiles with various line-of-sight velocities and/or widths caused by variations in the velocity or temperature from emission sources along the line of sight can result in asymmetries (e.g. Martínez-Sykora et al. 2011).

⋆ E-mail: razie.hosseini13@gmail.com (RH); safari@znu.ac.ir (HS)

Hara et al. (2008) used the extreme ultraviolet imaging spectrograph (EIS) observations and found blueward line asymmetries in the ARs. The same AR observation was systematically diagnosed by Peter (2010). They mentioned various physical processes (e.g. magnetic reconnection, plasma flows, asymmetric distribution of ions, i.e. non-Maxwellian distribution, heating, mass transfer, and instrument effect) responsible for asymmetric profiles in the ARs. The line asymmetries due to the outflow of plasmas have been studied in several attempts (Sakao et al. 2007; Del Zanna 2008; Harra et al. 2008; Brooks & Warren 2009; Ko et al. 2009; Kayshap et al. 2018a, 2021; Mishra et al. 2023). However, most favorably, the asymmetries in line profiles may indicate small-scale magnetic reconnections (Klimchuk 2006; Patsourakos & Klimchuk 2006).

Like, Peter (2010) investigated asymmetries in ARs' line profiles. So far, no work has been dedicated to diagnosing the asymmetries in QS, CH, and BPs using Interface Region Imaging Spectrometer (IRIS) high-resolution spectroscopic observations. Here, we investigate the Si IV line profile asymmetries in the QS, CH, and BPs. In addition, we use co-spatial-temporal (COST) observations from Atmospheric Imaging Assembly (AIA) and Helioseismic and Magnetic Imager (HMI) magnetograms.

Section 2 describes the used data and its analysis. In contrast, Section 3 gives the results and corresponding discussion. The last section (i.e. Section 4) concludes the main findings of this work.

## 2 OBSERVATION AND DATA ANALYSIS

We use the observations from the Interface Region Imaging Spectrograph (IRIS; De Pontieu et al. 2014), AIA (Lemen et al. 2012), and HMI (Scherrer et al. 2012). AIA and HMI instruments are onboard Solar Dynamics Observatory (SDO; Lemen et al. 2012).

IRIS provides high-resolution spectroscopic and imaging observations (i.e. slit-jaw images; SJI) of various solar atmosphere regions. IRIS has different SJI filters [e.g. NUV (2796 Å) and FUV (1330 Å)] with a pixel size of 0.16 arcsec and a cadence of ≈63 s which provide the images of the Sun of various layers from the photosphere to the transition region (TR).

IRIS collects spectra in three different bands of wavelengths, including 1332–1358 Å (FUV 1), 1389–1407 Å (FUV 2), and 2783–2851 Å (NUV). The effective spectral resolutions of IRIS are 26 mÅ for FUV 1/FUV 2, and 53 mÅ for NUV, respectively (De Pontieu et al. 2014). These have a pixel size of ≈0.16 arcsec along the slit, also the effective spatial resolution of the far UV spectra and near UV spectra ≈0.33 and ≈0.4 arcsec across the field of view (FOV), respectively. The time intervals between consecutive slit positions are ≈30 s. The FUV 1 band of IRIS contains two important spectral lines of TR, i.e. Si IV 1393.75 Å and 1402.77 Å. The Si IV 1393.75 Å is a strong line as the oscillator strength of this line is double of the Si IV 1402.77 Å. Hence, in the present analysis, We used the strong Si IV 1393.75 Å to investigate the line asymmetry in QS, CH, and BPs.

AIA uses eight separate passband (e.g. 1600 Å, 1700 Å, 304 Å, 171 Å, 193 Å, 211 Å, 335 Å, 94 Å) sensitive to plasma at various temperatures to investigate the Sun's atmosphere in the ultraviolet (UV) and extreme ultraviolet (EUV) region. The AIA images have a pixel size of ≈0.6 arcsec (Lemen et al. 2012).

HMI provides various observations, for instance, line-of-sight magnetogram, magnetic field inclination, continuum images, Dopp-lergrams, etc. Here, we used line-of-sight magnetograms with a cadence of ≈45 s and a pixel size of 0.5 arcsec. The present work utilizes three observations for QS and CH, simultaneously observed by IRIS, AIA, and HMI.

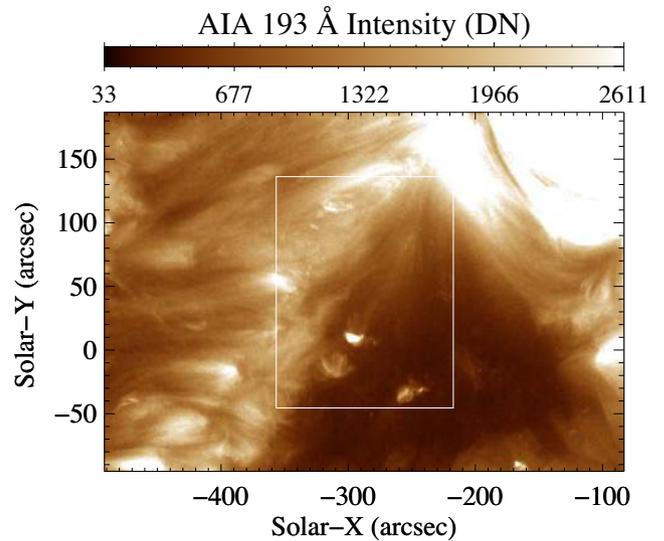

**Figure 1.** AIA 193 Å map for the first observation, i.e. data set 1. The overplotted white rectangular box shows the FOV of the IRIS raster. The auxiliary information about the data is in Table 2.

Table 2 tabulates the date, position ($X_{cen}$, $Y_{cen}$), FOV, and heliographic longitude ($\mu = \cos\theta$) for three observations (i.e. data sets 1, 2, and 3).

As we know, CHs are easily distinguishable from the QS at the coronal temperature, and the AIA 193 Å filter captures the emission from the solar corona (i.e. at high temperatures). Therefore, we utilize 193 Å images to discriminate the CH and QS boundaries. We used coordinated AIA data cubes and corresponding cutout HMI to display coronal features and magnetic structure in the FOV, respectively.

First of all, we would like to mention that the IRIS observation[1] was observed in large dense raster mode, i.e. the whole region was covered in 400 steps from 11:39:31 UT to 15:09:59 UT on 2015 April 26. Fig. 1 shows the AIA 193 Å intensity map. The over-plotted white box on the intensity map shows the FOV of the IRIS raster. Similar figures for data set 2 and data set 3 are displayed in Figs A1 and B1, respectively. First of all, we have produced the COST image from AIA 1600 Å image corresponding to IRIS rater (Si IV 1393.75 Å). To do so, we have followed the following steps.

(i) We first select the AIA 1600 Å image that is closest to the time of a particular IRIS slit. For example, in the case of data set 1, the first IRIS slit time is 11:40:03 UT. Here, it must be noted originally, it was 11:39:31 UT, but we have applied a binning of 2 × 2 in the x-direction and y-direction therefore after binning, the time of the first IRIS slit is 11:40:03 UT. Further, we take AIA 1600 Å of time $t$ = 11:39:51 UT. Here, we can see first IRIS slit time and AIA 1600 Å are very close to each other as the time difference is only 11 s.

(ii) Secondly, by applying the index2map.pro routine, we made the AIA 1600 Å map of the selected time frame (i.e. AIA 1600 Å image of t = 11:39:51 UT), and we also have the IRIS raster map by using make_map.pro. We know the x-coordinate of the first IRIS slit.

---

For example, in the case of data set 1, the x-coordinate of the IRIS first slit is −356.62 arcsec.

(iii) Now, using this location (i.e. −356.41 arcsec), we locate the nearest position (pixel) in the AIA 1600 Å map. The pixel number in the AIA 1600 Å image is 225. Further, we know the y-arcsec values for the first IRIS slit, and it is from −45.47 arcsec (lower y-arcsec) to 136.18 arcsec (upper y-arcsec). Then, we locate the position (pixel) corresponding to the lower y-arcsec in the AIA 1600 Å map. Similarly, we did locate the position (pixel) corresponding to the upper y-arcsec in the AIA 1600 Å map. The lower and upper y-pixels in the AIA 1600 Å image are 83 and 385 respectively.

(iv) In this way, for the first slit (i.e. one x-position), we take all the pixels along the y-direction of the AIA map, i.e. we took all y-pixels from 83 to 385 for one x-pixel of 225.

(v) Next, we repeated above mentioned steps for all 200 IRIS slits (i.e. originally, it was 400 steps, but we have applied a binning of 2 × 2 in both directions). Now, through this way, we produced the COST AIA 1600 Å image with the size of 200 × 302. Here, it should be noted that the size of the IRIS raster is 200 × 547. This is because the IRIS resolution is different than AIA. Further, using congrid.pro, we resized the y-dimension of COST AIA 1600 Å image to make it consistent with IRIS slit size along the solar y-direction.

(vi) Through the same process, we have produced the COST IRIS/SJI 1330 Å image from IRIS/SJI 1330 Å images, a COST AIA 193 Å image from AIA 193 Å images, and a COST HMI magnetogram from HMI magnetograms.

(vii) IRIS/SJI 1330 Å and AIA 1600 Å capture the emission from the same layer of the solar atmosphere, which makes them ideal for the alignment of the two instruments. Therefore, using get_correl_offset.pro, the alignment offset was calculated using COST AIA 1600 Å image and COST IRIS/SJI 1330 Å image. The offset is found 0.2 pixels in the x-direction and 1.0 pixels in the y-direction. We have applied the same offset to COST AIA 193 Å, COST 1600 Å image, and COST HMI magnetogram to make them consistent with the IRIS spectral raster.

(viii) We applied the above-mentioned procedures to all three data sets. Now, we can determine the boundaries of QS, CH, and BPs in the IRIS raster map that are recognized from COST AIA 193 Å image. Also, we can identify the position of asymmetric profiles in the COST HMI magnetogram and IRIS/SJI 1330 Å image that are determined from the IRIS raster. Finally, to show a perfect view, we made the map of the COST AIA 193 Å, COST AIA 1600 Å, COST IRIS/SJI 1330 Å image, and COST HMI magnetogram in the scale of IRIS raster.

Fig. 2 shows IRIS raster Si IV 1393.75 Åmap (panel a), COST AIA 193 Å map (panel b), COST HMI map (panel c), COST AIA 1600 Å map (panel d), and COST IRIS/SJI 1330 Å map (panel f) for data set 1. The same figures are shown in Figs A2 and B2 for data set 2 and data set 3, respectively.

We observe some dark and bright areas in the COST AIA 193 Å map, which are CH (dark area) and the QS regions (bright regions). Since the magnetic field is almost the same in QS and CH at the photosphere (Fig. 2c), the direct visual inspection identifying the CH and QS in the HMI (magnetograms) has yet to be reported. The HMI instrument measures the magnetic field at the solar photosphere. In addition to the dark (CH) and bright areas in COST AIA 193 Å intensity map (Fig. 2b), we also observe some compact, bright regions in COST AIA 193 Å map, and these regions are known the BPs. Here, please note all these regions (i.e. QS, CH, and BPs) are present in data set 2 (Fig. A2) and data set 3 (Fig. B2). Here, please note that bright and dark regions (as per COST AIA 193 Å) in COST AIA 1600 Å and IRIS/SJI 1330 Å maps have a similar amount of emission because both filters capture the emission from lower solar atmosphere. And, we know QS and CH have similar appearances in the lower solar atmosphere.

# 3 RESULTS AND DISCUSSION

## 3.1 Identification of QS, CH, and BPs

We can use various methods to draw the boundary between QS and CH. In a remarkable work, (Linker et al. 2021) discussed the various such methods, for instance, simple thresholding (THR; Rotter et al. 2012), SPoCA (Verbeeck et al. 2014), PSI-SYNCH and PSI-SYNOPTIC (Caplan, Downs & Linker 2016), CHIMERA (Garton, Gallagher & Murray 2018), and some more methods. Earlier, Arish et al. (2016) have also described some more methods to detect the CHs. Recently, Kayshap & Young (2023) used SPoCA to draw the boundary between QS and CH. In the present work, we used the simple thresholding approach to draw the boundary between QS and CH. The 35 per cent of the median coronal intensity is an appropriate intensity threshold to identify QS and CH (Hofmeister et al. 2017; Heinemann et al. 2018). Hence, we also apply the same intensity threshold (i.e. 35 per cent of the median coronal intensity) to draw the boundary of QS and CH.

For this purpose, we used four different AIA 193 Å full disc images on 2015 April 26. We calculate the median intensity (MI) for each AIA 193 Å full-disc image. The MI and 35 per cent of MI for each full disc image of 193 Å are tabulated in Table 1. Finally, we obtained the average of four intensity thresholds (i.e. 35 per cent medians), which is ∼102 DN.

In Fig. 2, we draw a contour (i.e. pink colour) to the intensity threshold (i.e. 102 DN) on the IRIS raster Si IV 1393.75 Åmap (panel a), COST AIA 193 Å map (panel b), COST HMI map (panel c), COST AIA 1600 Å map (panel d), and COST IRIS/SJI 1330 Å map (panel e).

We manually (visual inspection) identify four BPs within the full FOV of data set 1 via the intensity thresholds of trials and errors. The intensity threshold of 105 DN was applied to outline a BP in the CH, i.e. BP1. The rest of the three BPs (i.e. BP 2, 3, and 4) lie in the QS, and we used the intensity threshold of 150, 200, and 410 to outline BP2, BP3, and BP4, respectively. All BPs are outlined in light blue colour. Hence, in conclusion, we classified a total of three different types of features in this observation, namely, CH (area below pink contour), QS (area above pink contour), and four BPs (compact regions enclosed by the light blue contours). QS, CH, and BPs are located in the data set 2 (see Fig. A2) and data set 3 (see Fig. B2).

## 3.2 Nature of Si IV 1393.75 Å line profiles: single and double-Gaussian

The Si IV 1393.75 Å lines form in the TR, and mostly TR lines are asymmetric, double, or multipeak profiles (e.g. Peter 2000, 2001; Peter, Gudiksen & Nordlund 2006; Kayshap et al. 2018b; Mishra et al. 2023). Not only the TR spectral lines but the coronal spectral lines also show the asymmetry in the active regions as shown by Peter (2010). Here, we also found that some Si IV 1393.75 Å line profiles departed from single Gaussian and are likely to be asymmetric/double Gaussian profiles.The Si IV 1393.75 Å spectral line profiles are fitted with the single and double Gaussian functions. Fig. 3(a) shows the blueward asymmetric profile, and the profile is fitted by a single Gaussian (see blue dashed curve). Next, the double Gaussian fit is applied to the same blueward profiles (panel b). The resultant profile is shown by a solid red line while two Gaussians are displayed by two green dashed curves. Similarly, we have shown a redward asymmetric profile with a single (panel c) and double Gaussian fit (panel d), a double peak profile with a single (panel e) and double Gaussian fit (panel f), and a multipeak profile with a single (panel g) and double Gaussian fit (panel h). The locations corresponding to these four profiles shown in Fig. 3 are shown by four green diamonds in Fig. 2. We have mentioned the spectroscopic parameters [i.e. peak intensity $I$ (DN), Doppler velocity $v_D$ (km s$^{-1}$), Gaussian width $\sigma$ (km s$^{-1}$)] and goodness-of-fit $\chi^2$ in each panel of Fig. 3. The right column shows the parameters for the components of the double Gaussian model, i.e. for two Gaussians.

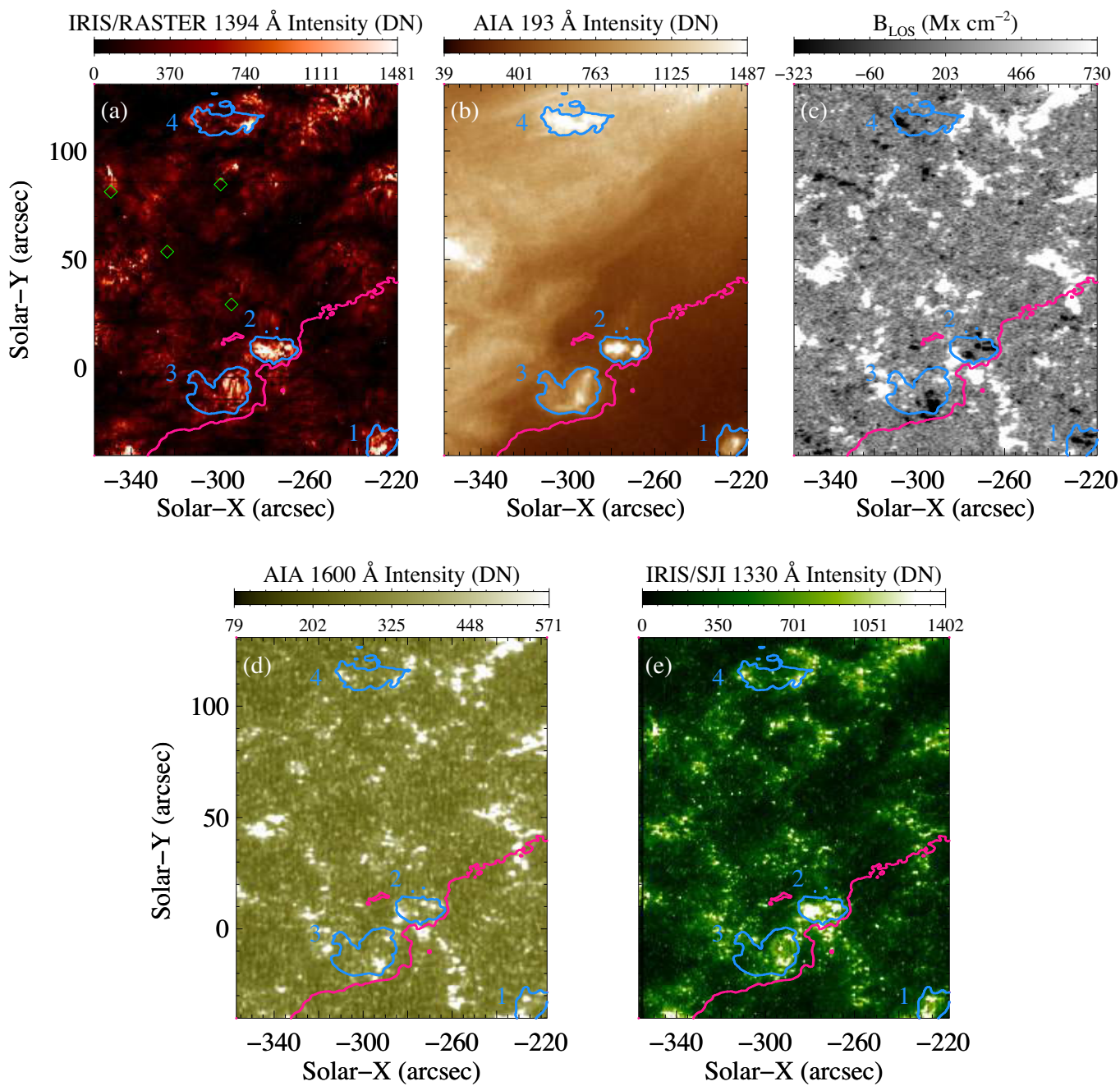

**Figure 2.** IRIS raster Si ɪᴠ 1393.75 Å map (panel a), COST AIA 193 Å map (panel b), COST HMI map (panel c), COST AIA 1600 Å map (panel d), and COST IRIS/SJI 1330 Å map (panel e) for the data set 1. The pink contour indicates the boundary of QS and CH. The light blue shows the border of BPs that mark with 1, 2, 3, and 4. The green diamonds indicate the position of four spectral lines shown in Fig. 3.

**Table 1.** The median intensity (MI) and its 35 per cent for AIA full-disc image at 193 Å.

| Sr. No. | Time of observation | MI | 35 per cent MI |
|---|---|---|---|
| 1 | 00:59:30 | 301.0 | 105.35 |
| 2 | 06:59:30 | 295.0 | 103.25 |
| 3 | 12:59:30 | 292.0 | 102.20 |
| 4 | 16:59:30 | 282.0 | 98.70 |

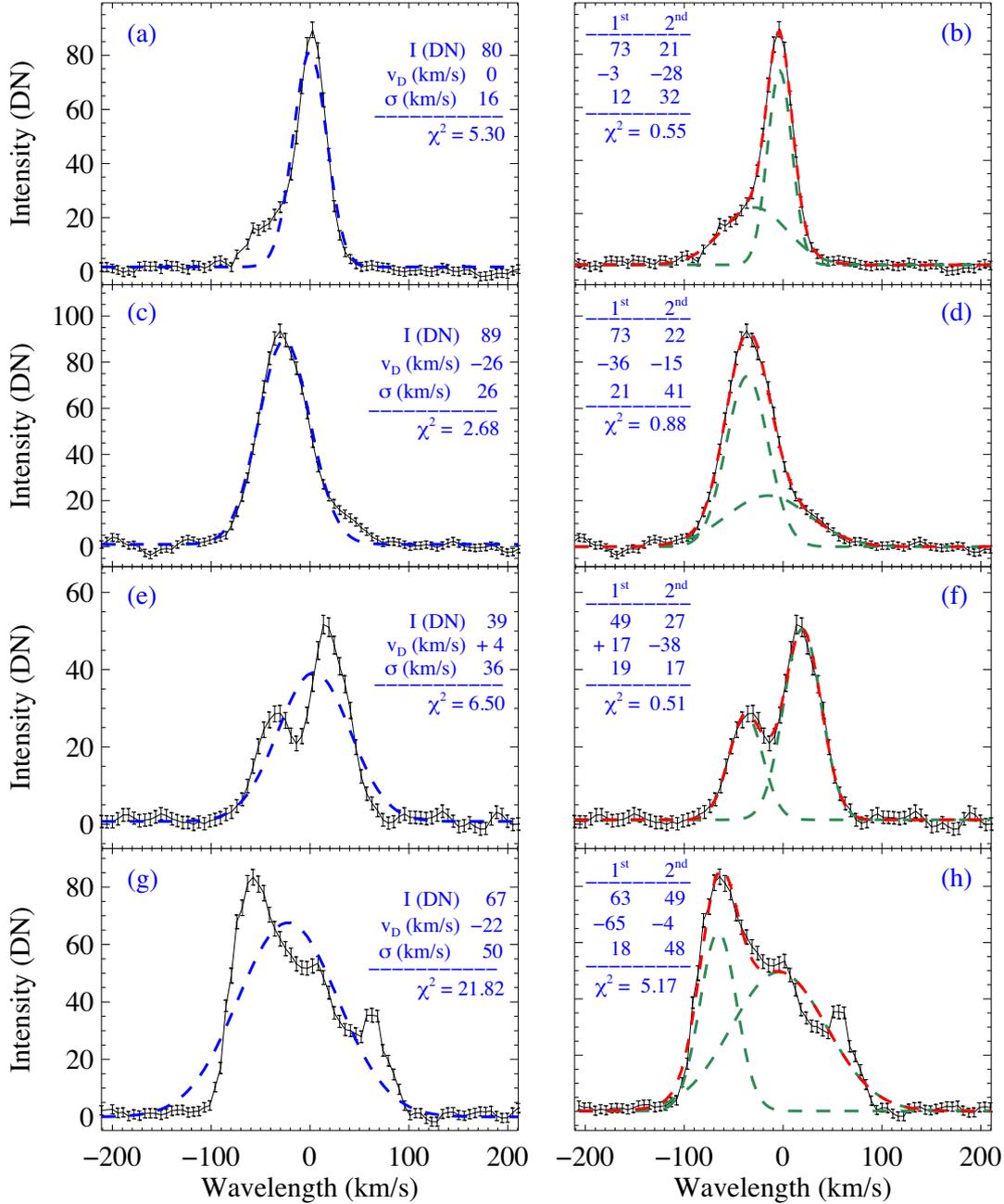

**Figure 3.** The blueward asymmetric profile (black solid curve) with single Gaussian fit (blue dashed curve; panel a) and double Gaussian fit (red dashed curve; panel b). In the same way, we have shown redward asymmetric profile (panels c and d), double peak profiles (panels e and f), and complex multipeak profile (panels g and h). The peak intensity $I$, Doppler velocity $v_D$, and width $\sigma$ are in DN, km s$^{-1}$ and km s$^{-1}$, respectively. The locations of these four spectral profiles are shown by the green diamonds in Fig. 2.

As the Fig. 3 shows, the single Gaussian largely deviates from the line profiles, so the $\chi^2$ in the left column is much greater than one. On the contrary, the double Gaussian fits very well as the $\chi^2$ is less than one (i.e. panels b, d, and f) except the bottom right panel (i.e. panel h). The last profile (i.e. panel h) is a complex multipeak profile. Please note that such complex and multipeak profiles are rare in our observations.

### 3.3 Statistical analysis of QS, CH, and BPs

Firstly, each Si IV 1393.75 Å spectral profile of the region of interest (ROI; as FOV of panels shown in Fig. 2) from the first observation

(2015 April 26) is fitted by single Gaussian. Through this approach, we estimate the peak intensity, centroid, and Gaussian sigma for each location (pixel) of ROI. The centroid is converted into the Doppler velocity using the rest wavelength. To estimate the rest wavelength of Si IV 1393.75 Å, we use the cool spectral line method as described by Peter & Judge (1999). We have used the cool line Fe II 1405.61 Å in the present analysis. We apply the following steps to determine the rest wavelength of the Si IV 1393.75 Å.

(i) First, we select a quiet area (a box of 10 × 10 pixels) in the QS. We average the spectra of the box to get averaged spectra.

(ii) We fitted a single Gaussian to the averaged spectra to determine the Gaussian's centroid of Fe II 1405.61 Å.

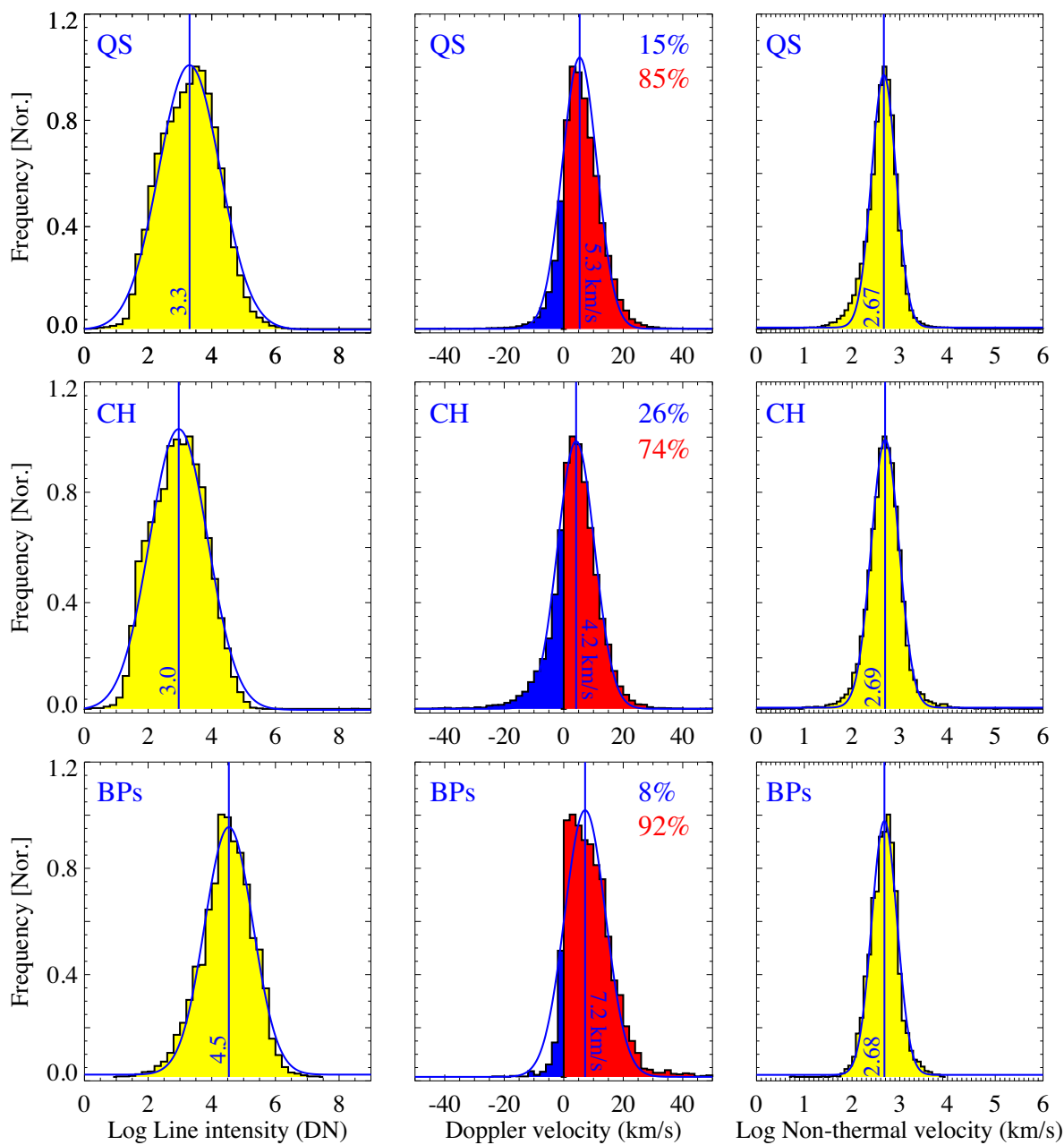

**Figure 4.** The peak intensity (logarithm), Doppler velocity, and non-thermal velocity (logarithm) histograms of QS (top row), CH (middle row), and BPs (bottom row), which are deduced using the single Gaussian fit on the data set 1. The blue and redshift contributions are indicated for each region in the histogram (middle column). Further, a Gaussian function is fitted to the histograms (see blue curve) to obtain the mean and standard deviation of the corresponding parameter. The solid blue vertical line in each histogram is at the mean value of the corresponding parameter. Further, the mean values are also mentioned in each panel.

(iii) The centroid is obtained to be 1405.59 Å, considering the rest wavelength of Fe II 1405.61 Å. The difference between the observed centroid and standard wavelength Fe II is 0.02, i.e. 1405.61−1405.59 = 0.02 Å.

(iv) Using 0.02 Å correction obtained by the above steps, the Si IV 1393.75 Å rest wavelength is estimated to be 1393.73 Å.

For any spectral line profile, the total line width is the sum of thermal, non-thermal, and instrumental line widths. We neglect the generally small natural width in this analysis. We fed the observed (1/e) line width ($\sigma$ of Gaussian fit in km s$^{-1}$) and instrumental FWHM in angstrom unit (26 mÅ for the FUV) to iris_nonthermalwidth.pro to obtain the non-thermal velocity in km s$^{-1}$. The non-thermal velocity shows the activity level in the *in-situ* plasma.

We have already identified QS, CH, and BPs (Section 3.2), and then we collected all spectroscopic parameters (i.e. intensity, Doppler velocity, and non-thermal velocity) for each region separately. We have produced histograms of these spectroscopic parameters using these values for QS, CH, and BPs. The top row of Fig. 4 shows the histogram of the logarithm (natural) of peak intensity (left column), Doppler velocity (middle column), and the logarithm of non-thermal velocity (right column) of QS (top row). Similarly, the middle and

**Table 2.** Summary of three sets of IRIS raster observations, including date, position of centre ($X_{cen}$, $Y_{cen}$), FOV, and heliographic longitude ($\mu = \cos\theta$).

| Set | Date time | ($X_{cen}$, $Y_{cen}$) | Raster FOV | $\mu$ |
|---|---|---|---|---|
| Data set 1 | 2015-04-26 11:39:31–15:09:56 | (−288 arcsec, 45 arcsec) | (141 arcsec, 174 arcsec) | 0.95 |
| Data set 2 | 2014-07-24 11:10:28–14:40:53 | (128 arcsec, −180 arcsec) | (141 arcsec, 174 arcsec) | 0.97 |
| Data set 3 | 2014-07-26 00:10:28–03:40:53 | (469 arcsec, −167 arcsec) | (141 arcsec, 174 arcsec) | 0.85 |

**Table 3.** The mean and standard deviation (obtained from histograms by the Gaussian fit) of the peak intensity (logarithm), Doppler velocity, and non-thermal velocity (logarithm) for line profiles of three data sets (Figs 4, A3, and B3).

| Region | Data | Log intensity (DN) | Doppler velocity (km s$^{-1}$) | Log non-thermal velocity (km s$^{-1}$) |
|---|---|---|---|---|
| QS | Data set 1 | 3.3 ± 0.9 | 5.3 ± 5.8 | 2.67 ± 0.25 |
| | Data set 2 | 3.1 ± 0.8 | 6.5 ± 6.1 | 2.61 ± 0.26 |
| | Data set 3 | 3.2 ± 0.9 | 5.5 ± 4.4 | 2.60 ± 0.26 |
| CH | Data set 1 | 3.0 ± 0.8 | 4.2 ± 6.2 | 2.69 ± 0.28 |
| | Data set 2 | 2.9 ± 0.7 | 5.8 ± 6.2 | 2.63 ± 0.27 |
| | Data set 3 | 3.1 ± 0.7 | 4.7 ± 5.7 | 2.62 ± 0.26 |
| BPs | Data set 1 | 4.5 ± 0.7 | 7.2 ± 6.8 | 2.68 ± 0.26 |
| | Data set 2 | 4.8 ± 0.9 | 8.9 ± 6.7 | 2.71 ± 0.22 |
| | Data set 3 | 3.8 ± 0.9 | 7.0 ± 8.0 | 2.63 ± 0.24 |

These parameters (peak intensity, Doppler velocity, and non-thermal velocity) were the outputs of the single Gaussian modeling of the line profiles.

bottom rows of Fig. 4 show the histogram of the logarithm (natural) of peak intensity (left column), Doppler velocity (middle column), and the logarithm of non-thermal velocity (right column) of CH and BPs.

Further, we applied the Gaussian fit onto each histogram of Fig. 4 (see blue solid curve in each panel), and the centroid of Gaussian given the mean value of the corresponding parameter. Further, we have drawn the vertical blue solid line at the mean value of each parameter. The mean of peak intensities (in logarithmic) are 3.3 ± 0.9, 3.0 ± 0.8, and 4.5 ± 0.7 for QS, CH, and BPs, respectively.

The BPs intensities are more than three times the intensities of QS and CH. Hence, we can peak intensities of BPs at the TR are much higher than QS and CH. On the other hand, the peak intensities of CH are slightly smaller than QS. While, at the coronal temperatures, the significantly lower intensity of CHs than QS/BPs is a very well-known fact (Fig. 2b, map of 193 Å). Here, we found that CH has a slightly lower intensity than QS within the TR. Similar findings (i.e. lower intensity of CH than QS in the TR) are previously reported by Tripathi, Nived & Solanki (2021). The similar range/behaviour of intensities are found for data set 2 (Fig. A2) and data set 3 (Fig. B2). The mean and standard deviations of peak intensity (logarithm), Doppler velocity, and non-thermal velocity (logarithm) from all three data sets are tabulated in Table 3.

The mean Doppler velocities are 5.3 ± 5.8, 4.2 ± 6.2, and 7.2 ± 6.8 km s$^{-1}$ in QS (top middle panel), CH (middle middle panel), and BPs (bottom middle panel), respectively. Hence, on average, all these regions are redshifted. Further, we have also noticed that the redshifted profiles are more frequent than the blueshifts in all three regions, i.e. 85 per cent, 74 per cent, and 92 per cent profiles are redshifted in QS (top middle panel), CH (middle middle panel), and BPs (bottom middle panel), respectively. While, only 15 per cent, 26 per cent, and 8 per cent profiles are blueshifted in QS, CH, and BPs. Similar findings about the Doppler velocity are found in data set 2 (see Fig. A3) and data set 3 (see Fig. B3). Finally, the mean Doppler velocities with the standard deviation of QS, CH, and BPs from all three observations are tabulated in Table 3. Hence, we can say that the redshifts dominate in the TR of QS, CH, and BPs. Similar findings are already reported (e.g. Peter & Judge 1999; Kayshap & Dwivedi

2017; Kayshap & Young 2023). The redshift in CH is slightly lower than the QS, while BPs have significantly higher values than QS and CH. This behaviour is consistent for all three data sets (Table 3, Figs A3 and B3).

The Doppler shifts in QS, CH, and AR show the centre-to-limb variations (CLV; e.g. Peter & Judge 1999; Kayshap & Young 2023; Rajhans et al. 2023), i.e. the Doppler velocity varies from its minimum (zero) at the solar limb to maximum value (few km s$^{-1}$) at the disc centre. Hence, the mean Doppler velocity of QS and CH depends on the locations at the solar disc. Peter & Judge (1999) obtained the QS redshift around 6.5 km s$^{-1}$ at the disc centre for Si IV profiles. Chae, Schühle & Lemaire (1998b) have also reported almost similar redshifts (i.e. 7.8 km s$^{-1}$) in Si IV lines at the disc centre. Recently, Kayshap & Young (2023) have also reported a Doppler velocity of around 6.0 km s$^{-1}$ in QS and 5.0 km s$^{-1}$ in CH near the disc centre. In the current work, the heliographic longitude ($\mu$) for QS and CH (see Table 2) are close to the disc centre (0.85–0.97). For data set 1, we obtain the mean Doppler shifts about 5.3 ± 5.8 and 4.2 ± 6.2 km s$^{-1}$ (Table 3) for QS and CH, respectively. Please note that almost similar values are reported for data set 2 and data set 3. Hence, mean Doppler shifts in QS and CH are consistent with the previous findings. Although, a significantly higher redshift (i.e. ∼10 km s$^{-1}$) in QS is reported by Teriaca, Banerjee & Doyle (1999) using Si IV 1393.75 Å.

It is interesting that standard deviations of Doppler velocity for single Gaussian fitting are in a similar order to the mean value. In other words, the standard deviation of Doppler velocity is sometimes even more significant than the mean value (see Table 3), indicating that a higher standard deviation means a greater spread of Doppler velocity. We know that the Doppler shift of any spectral line depends on the activity level that exists at a particular height where the spectral line has formed. So, the more variations in the Doppler velocity, as we have seen in the present case, justify that a large scale of the released energy (i.e. significant variations in the released energy) is responsible for the shift.

Finally, the right column of Fig. 4 displays the histogram of the logarithm of non-thermal velocity for QS, CH, and BPs from the first observations. The mean non-thermal velocities (logarithm) are almost the same ($\sim$15 km s$^{-1}$) for all three regions (Table 3) for data set 1. The blue solid vertical line in all panels of the right column represents the mean value of non-thermal velocity. An almost similar range of non-thermal velocities (i.e. around 15 km s$^{-1}$) is reported in the other two observations also. Similar to Doppler velocity, the non-thermal velocity in QS and CH also show CLV (Rao, Del Zanna & Mason 2022; Kayshap & Young 2023), i.e. the non-thermal velocity varies as per the $\mu$ angle. The non-thermal velocity is maximum at the solar limb (i.e. $\mu = 0.0$) and minimum at the disc centre (i.e. $\mu = 1.0$).

**Table 4.** The number and percentage of asymmetric profiles for each region of data sets 1, 2, and 3.

| Region | Data | Total profiles | Asymmetric profiles | Percentage |
|---|---|---|---|---|
| QS | Data set 1 | 79 147 | 6857 | 8.7 per cent |
|  | Data set 2 | 55 352 | 5922 | 10.7 per cent |
|  | Data set 3 | 66 764 | 7577 | 11.3 per cent |
| CH | Data set 1 | 19 390 | 1577 | 8.1 per cent |
|  | Data set 2 | 46 837 | 4665 | 10 per cent |
|  | Data set 3 | 32 621 | 3447 | 10.6 per cent |
| BPs | Data set 1 | 4463 | 1104 | 24.7 per cent |
|  | Data set 2 | 811 | 264 | 32.6 per cent |
|  | Data set 3 | 3615 | 599 | 16.6 per cent |

Using Solar Ultraviolet Measurements Of Emitted Radiation (SUMER)/Heliospheric Observatory (SOHO) observations, Chae, Schühle & Lemaire (1998b) obtained the non-thermal velocity around 17–18 km s$^{-1}$ at the log T [K] = 4.90 (formation temperature of Si IV). Recently, Kayshap & Young (2023) also shown that non-thermal velocity is also around 15–16 km s$^{-1}$ near the disc centre [see figs 3 and 5 of Kayshap & Young (2023)]. The used observations in this work are located near the disc centre (see Table 2). Hence, the reported values of non-thermal velocity match the previous values. However, Teriaca, Banerjee & Doyle (1999) have shown significantly higher values of QS non-thermal velocity (i.e. $\sim$ 26 km s$^{-1}$) in the same Si IV 1393.755 Å spectral line at the disc centre.

As we already mentioned, not all the spectral profiles are single Gaussian profiles, but some are asymmetric/double peak profiles (see Section 3.2). We applied two conditions to qualify any spectral profiles as asymmetric profiles, namely, (1) the asymmetric profile must have a goodness-of-fit greater than one for the single Gaussian fits, and (2) the peak intensity of the second component must be greater than 0.25 of the first component's peak intensity. We ignored the weak asymmetry profiles (i.e. profiles with the second component peak weaker than 0.25 times of first component peak intensity) in this analysis. Table 4 displays the statistics of asymmetric line profiles for QS, CH, and BPs of all three data sets. We obtain that about 8–11 per cent line profiles of QS and CH satisfy the above condition of the asymmetric. Meanwhile, 17–33 per cent line profiles of BPs contain secondary component that satisfy the above criteria.

Next, we investigated the locations of asymmetric profiles in QS, CH, and BPs. To do so, we have overplotted asymmetric profile locations on COST IRIS/SJI 1330 Å intensity map (Fig. 5a), COST HMI magnetogram (Fig. 5b), and COST AIA 193 Å intensity map (Fig. 5c). The similar behavior was observed for data sets 2 (Fig. A4) and 3 (Fig. B4). Most asymmetric profiles occurred in the large-scale bright lane-like regions, as seen in the COST IRIS/SJI 1330 Å (Fig. 5a). The large-scale bright lane-like regions in the COST IRIS/SJI 1330 Å map are the network regions (e.g. Chen et al. 2019). Further, the networks correspond to magnetic flux concentration in the COST HMI map. Next, the boundary between QS and CH is also the position of many asymmetric profiles. Here it should be noted that the boundary between QS and CH is a complex magnetic field environment, i.e. close loops of QS interact with the open magnetic fields of CHs (Madjarska et al. 2003; Bale et al. 2023). In addition to the network and boundary between QS and CH, we also observed numerous locations within BPs corresponding to the asymmetric line profiles. The BPs are complex magnetic environments and are composed of the loop-like structure (Kayshap & Dwivedi 2017; Madjarska 2019; Bale et al. 2023). Hence, finally, we mention that occurring asymmetric profiles above the concentrated complex magnetic field region indicate the magnetic source (e.g by magnetic reconnection) behind the formation of these asymmetries.

We again take the spectra from the selected asymmetric locations, and the profiles are fitted by the double Gaussian function. Now, we estimated the Doppler velocity and non-thermal velocity for the secondary component. Finally, Fig. 6 shows the histograms of the Doppler velocity (left column), and the logarithm of non-thermal velocity (right column) of the secondary component of QS (top row), CH (middle row), and BPs (bottom row) of the data set 1. In the case of a single Gaussian fit, the majority of the Si IV 1393.75 Å dominated by the redshifts (see Fig. 4). However, statistics of the Doppler velocity of the secondary component of Si IV 1393.75 Å are significantly different. For QS, less than half of the locations (i.e. 44 per cent) are blueshifted, while more than 56 per cent are redshifted. However, in the case of CH, more than half of the locations are blueshifted (i.e. 51 per cent), and less than half of the locations (i.e. 49 per cent) are redshifts. Contrary to QS and CH, about 71 per cent of the secondary components of BPs show redshifts, and the remaining 29 per cent are blueshifted. Similar to Fig. 4, the histograms are fitted with Gaussian (see blue solid curve in each panel of Fig. 6), and estimate the mean value of the corresponding parameter (see solid blue vertical line in each panel of Fig. 6). On average (i.e. the mean value of the secondary component of QS and CH have almost zero Doppler shift, while the secondary component of BPs shows redshifts of around 4.0 km s$^{-1}$.

The histogram for the logarithm (natural) non-thermal velocity of the secondary component of BPs has a peak at 3.14 ± 0.35, which is significantly higher than the mean of the non-thermal velocity of the secondary component of QS (2.96 ± 0.45) and CH (2.82 ± 0.62). The similar behaviour of non-thermal velocity of the secondary components of asymmetric profiles for data sets 2 and 3 are shown in Figs A5 and B5, respectively. Hence, the non-thermal velocity of secondary Gaussian is significantly higher in BPs than QS and CH. Further, Table 5 tabulates the mean and standard deviation values of Doppler velocity and non-thermal velocities (logarithm) of the secondary component of QS, CH, and BPs in all three observations.

We must mention that the asymmetric profiles are ubiquitous in the TR as reported previously (e.g. Dere & Mason 1993; Peter 2000, 2001; Tian et al. 2014a; Kayshap et al. 2018b, 2021; Kayshap &

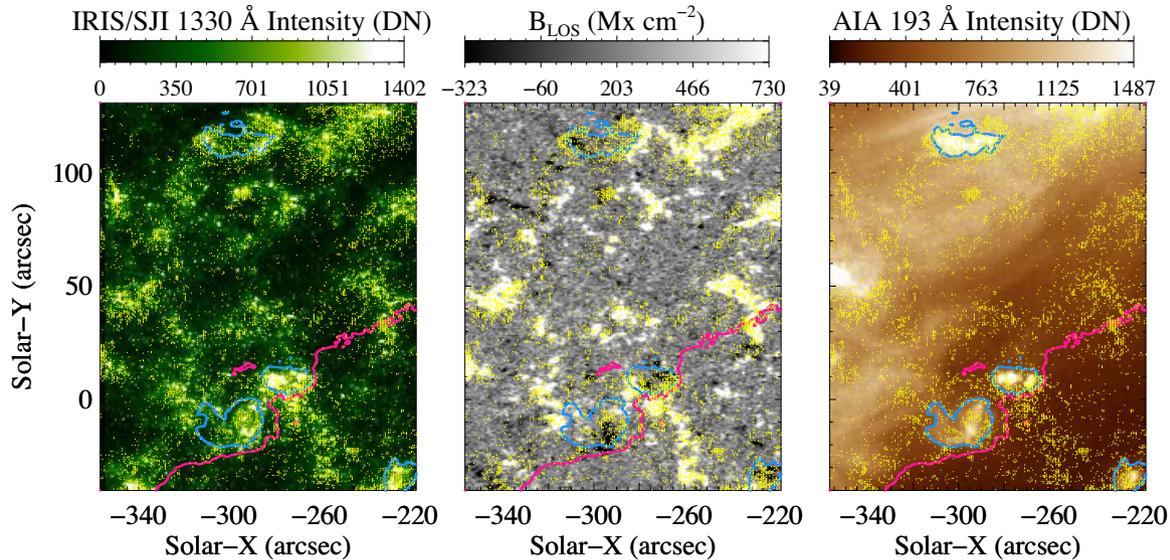

**Figure 5.** The yellow dots show the position of asymmetric profiles on the COST IRIS/SJI 1330 Å (left panel), COST HMI (middle panel), and COST AIA 193 Å (right panel) maps. The pink contour demonstrates the QS and CH boundary. The light blue contours indicate the border of BPs.

Young [2023]). Furthermore, the TR spectral lines and the coronal spectral lines also show asymmetry. A secondary physical process on top of the background physical process leads to the formation of asymmetric profiles. Such physical processes can be explosive events (Dere & Mason [1993]), shocks in the sunspot umbra (Tian et al. [2014b]; Kayshap et al. [2021]), magnetic reconnection (Peter [2001]), and rotating motion in solar jets (Kayshap et al. [2018a]; Mishra et al. [2023]). In a remarkable work, Peter ([2010]) explained many physical processes that can produce asymmetric profiles within high magnetic field regions (i.e. ARs). Finally, we can say that such physical processes can insert another component in the Gaussian profiles and the profile becomes asymmetric profiles.

We have already mentioned that most of the asymmetric profiles exist in the vicinity of the network regions, the boundary between QS and CH, and of course, BPs. The networks are the most preferable sites of small-scale jets (network jets), and they are most probably produced by small-scale magnetic reconnection (Tian et al. [2014a]; Kayshap et al. [2018a]). Further, we must mention that network jets are prevalent in networks, i.e. the prevalence of magnetic reconnection in the vicinity of networks (Tian et al. [2014a]; Narang et al. [2016]; Kayshap et al. [2018a]). Recently, Kayshap et al. ([2018a]) reported the occurrence of asymmetric profiles within the network jets due to the rotating plasma which is itself a result of magnetic reconnection. Further, the boundary between QS and CH is again a preferable site of the meeting of closed-loop and open magnetic field, i.e. it is the most probable site of the magnetic reconnections (e.g. Madjarska & Wiegelmann [2009]; Subramanian, Madjarska & Doyle [2010]; Huang et al. [2012]; Madjarska et al. [2012]; Chen et al. [2019]; Upendran & Tripathi [2022]). At last, after the two most probable sites of asymmetric profiles (i.e. network regions and the boundary between QS and CH), the third region (i.e. BPs) is also the most probable site of the small-scale magnetic reconnection (e.g. Kayshap & Dwivedi [2017]; Madjarska [2019]). Finally, we say that asymmetric profile-dominated network regions are very much prone to the magnetic reconnection (Porter & Dere [1991]; Dere [1994]; Chae et al. [1998a]; Innes & Tóth [1999]; Roussev et al. [2001a], [b]; Roussev & Galsgaard [2002]; Chen & Priest [2006]; Innes et al. [2015]).

The height of magnetic reconnection is an important parameter to reveal the nature of plasma flows in any particular layer of the solar atmosphere. The plasma will flow in the up direction above the reconnection site, i.e. the secondary Gaussian will be blueshifted. While below the reconnection site, the plasma will be in the down direction, and the profile (or secondary Gaussian) will be redshifted. Moreover, most importantly, the profile (or secondary Gaussian) will show zero shift at the reconnection site. In various works, several spectral lines, forming at different heights of the solar atmosphere, are used to investigate plasma flow (i.e. Doppler shifts that are deduced using single Gaussian fit) from the photosphere to the corona in QS, CH, and ARs (e.g. Peter & Judge [1999]; Teriaca, Banerjee & Doyle [1999]; Dadashi, Teriaca & Solanki [2011]; Kayshap, Banerjee & Srivastava [2015]). Such research works reported that TR spectral lines are redshifted and the coronal lines are blueshifted. It means the reconnection is happening somewhere in between TR and corona, i.e. around log T [K] = 5.7 in QS and CH (Peter & Judge [1999]; Teriaca, Banerjee & Doyle [1999]; Kayshap, Banerjee & Srivastava [2015]). Therefore, it might imply that the TR lines forming below the reconnection site are redshifted, while coronal lines forming above the reconnection site are blueshifted. These works show that the Si IV is redshifted by ∼6–7 km s⁻¹ at the disc centre. In the case of a single Gaussian fit, we also found a similar range of Doppler velocities for Si IV (see Fig. 4 and Table 3).

However, the most frequent Doppler shift of secondary Gaussian is almost zero for QS and CH at a temperature of log T [K] = 4.85, i.e. the formation temperature of Si IV 1393.75 Å. For the first time, we have reported that, most probably, the secondary Gaussian originates at the formation height of the Si IV line in QS and CH. In the case of BPs, the mean Doppler velocity of the secondary Gaussian is ∼4.0 km s⁻¹, and most of the secondary Gaussian (i.e. 71 per cent) are redshifted. Hence, contrary to QS and CH, the magnetic reconnection in BPs, which is responsible for secondary Gaussian, occurs at a height well above the formation height of Si IV. However, the secondary components of some other spectral lines (i.e. forming above and below the Si IV line) must be analysed to draw a firm conclusion.

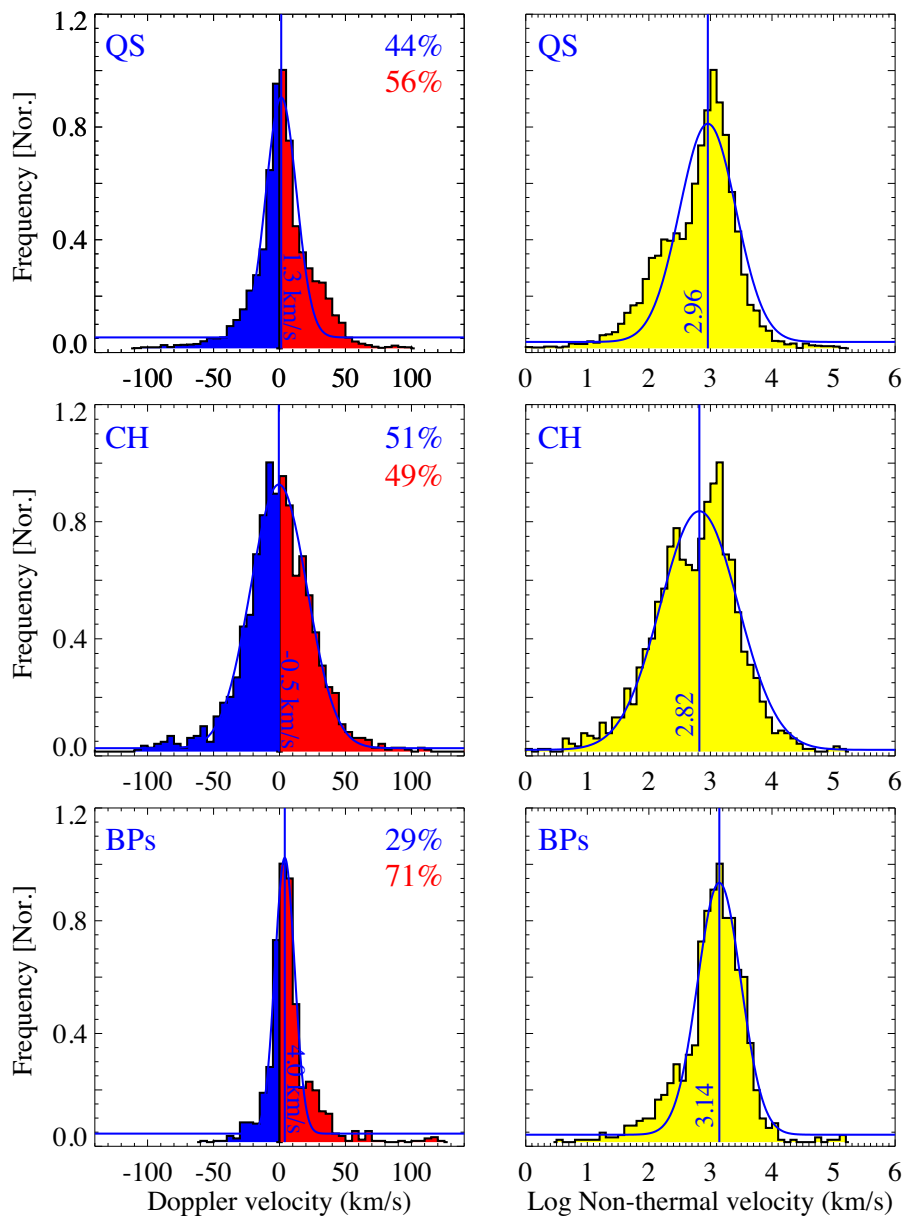

**Figure 6.** Histogram of the Doppler velocity, and non-thermal velocity (logarithm) of the secondary component of asymmetric profiles for QS (top row), CH (middle row), and BPs (bottom row) for data set 1. The blue and redshift contributions are indicated for each histogram (left column). A Gaussian function is fitted to the histograms (see blue curve) to obtain the mean and standard deviation, and we have written the mean values of the corresponding parameter (solid blue vertical line).

Similar to the single Gaussian fit results (Table 3), the standard deviation of the Doppler velocity of the secondary Gaussian is also very high (see Table 5). We would even say that the standard deviation of the Doppler velocity of the secondary Gaussian is higher than the standard deviation of the Doppler velocity of the single Gaussian. Hence, we can say that the range of released energy, which is responsible for the secondary Gaussian in the spectral profile, is broader than the released energy range, which is responsible for the shift of the primary component of the spectral profile.

It is found the non-thermal velocities of the secondary are almost the same in QS and CH. The mean non-thermal velocity (Fig. 6) is approximately 18.0 km s$^{-1}$ in QS and CH, while for BPs is ∼ 23.0 km s$^{-1}$. Hence, the activity level is high in the BPs in comparison to the QS and CH. Further, this is the first time we measured the on-average behaviour of Doppler velocity and non-thermal velocity of the secondary component in QS, CH, and BPs.

Fig. 7 displays the probability density function of peak intensity for the second component of asymmetric profiles of QS (left panel), CH (middle panel), and BPs (right panel) for the first data set. As shown in the figure, the peak intensities of QS and BPs are slightly greater than the peak intensity of CHs. We observe the power-law-like behaviour at the tails of the distribution of the peak intensity of the second component of QS, CH, and BPs for data set 1. The similar behaviour for the peak intensity of the second component (i.e. power-law-like behaviour) for data sets 2 and 3 are represented in Figs A6 and B6, respectively. Heavy tail distributions, such as

**Table 5.** The mean and standard deviation (obtained from histograms by the Gaussian fit) of the Doppler velocity and non-thermal velocity (logarithm) for line profiles of three data sets (Figs 6, A5, and B5).

| Region | Data | Doppler velocity (km s$^{-1}$) | Log non-thermal velocity (km s$^{-1}$) |
|---|---|---|---|
| QS | Data set 1 | 1.3 ± 11.6 | 2.96 ± 0.45 |
|  | Data set 2 | 5.9 ± 20 | 2.76 ± 0.6 |
|  | Data set 3 | 3.6 ± 14.7 | 2.91 ± 0.53 |
| CH | Data set 1 | −0.5 ± 21.4 | 2.82 ± 0.62 |
|  | Data set 2 | 1.9 ± 22.2 | 2.84 ± 0.63 |
|  | Data set 3 | 1.3 ± 19.3 | 2.89 ± 0.63 |
| BPs | Data set 1 | 4.0 ± 7.4 | 3.14 ± 0.35 |
|  | Data set 2 | 7 ± 7.4 | 3.07 ± 0.19 |
|  | Data set 3 | 3.9 ± 7.7 | 3.03 ± 0.29 |

These parameters (Doppler velocity and non-thermal velocity) were the outputs of the double Gaussian modeling of the line profiles.

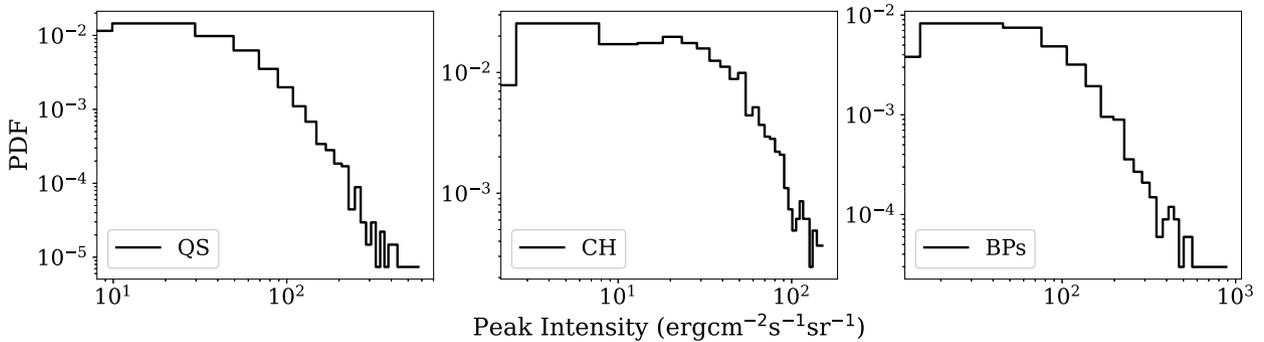

**Figure 7.** Probability density function of peak intensity for the second component of asymmetric profiles of QS (left panel), CH (middle panel), and BPs (right panel) for data set 1.

power laws, are features of systems controlled by complex generative mechanisms (Bak, Tang & Wiesenfeld 1987; Broder et al. 2000; Pisarenko & Sornette 2003; Newman 2005; Clauset, Rohilla Shalizi & Newman 2009; Klaus, Yu & Plenz 2011; Marković & Gros 2014). Solar flare energies and solar small-scale brightenings distributions are examples of power-law scaling behaviour (Aschwanden et al. 2016; Berghmans et al. 2021; Alipour et al. 2022). The power law-like distribution for the peak intensity of the secondary component of line profiles at TR may be the signature of self-similar or self-organized systems. The preferential attachment or automaton avalanche is among the generative mechanisms that produce the power law distribution for events. The power-law-like behaviour of flaring-like features (e.g. solar flares and small-scale events) is related to the magnetic reconnection (Parker 1983; Lin et al. 1984; Hudson 1991; Georgoulis, Vilmer & Crosby 2001; Klimchuk & Cargill 2001; Aschwanden et al. 2016; Farhang, Safari & Wheatland 2018; Tripathi, Nived & Solanki 2021). We have already justified that magnetic reconnection is the most favorable mechanism for the origin of the secondary component. Moreover, the power law behaviour again indicates that magnetic reconnection may generate the secondary component in all three regions, i.e. QS, CH, and BPs.

## 4 CONCLUSIONS

Here, using three different IRIS observations (Table 2), we investigated the characteristics of Si IV spectral lines in QS, CH, and BPs. The principal analysis of the present work was based on the single and double Gaussian fitting of spectral line profiles to obtain the physical parameters. All three regions (i.e. QS, CH, and BPs) in all three observations have a significant percentage of asymmetric profiles and a tiny percentage of very complex profiles with multiple peaks. From this analysis, we find some significant results that can improve our understanding of QS, CH, and BPs, which are concluded below.

(i) Most asymmetric profiles reside in the complex magnetic field environment (i.e. magnetic reconnection prone region), namely, networks, the boundary between QS and CH, and BPs.

(ii) The magnetic reconnection is likely responsible for the production of asymmetric profiles as the asymmetric profiles originate within the magnetic reconnection prone regions. In addition, the intensity of the secondary Gaussian component (in all regions) follows power-law behaviour, which justifies the occurrence of magnetic reconnection in these regions.

(iii) For the first time, we performed a systematic statistical investigation of the secondary Gaussian component in the QS, CH, and BPs, and we found an on-average zero Doppler shift of the secondary component in all three regions.

(iv) The statistical analysis reveals that the mean Doppler shift of the secondary component of BPs is significantly redshifted, i.e. around 4.0 km s$^{-1}$.

(v) We find that the non-thermal velocities of the secondary components of QS and CH are very similar to each other (around ∼18 km s$^{-1}$). In contrast, the non-thermal velocity of the secondary Gaussian of the BPs is significantly higher than QS or CH. Hence, unresolved motions (i.e. non-thermal motions) in BPs are more than QS and CH.

(vi) The standard deviations of the Doppler velocity in both single Gaussian (Table 3) and secondary Gaussian (Table 5) are significantly large. It justifies that the range of released energy, responsible for the Doppler shift of the primary or secondary component of the spectral profile, through the magnetic reconnection process in the TR is extensive.

Lastly, we mention that the primary Gaussian of the TR spectral line (Si iv) is redshifted, i.e. most of the profiles (i.e. more than 75 per cent) are redshifted in all types of regions. However, the secondary Gaussian has a different nature than the primary Gaussian. The secondary Gaussian in more than half of the locations of QS and CH is blueshifted and red-shifted at the rest of the locations. Hence, this secondary Gaussian shows that more upflows than downflows exist in the TR of QS and CH. The magnetic reconnection that probably produces the secondary component occurs at the formation height of Si iv as the mean Doppler velocity of the secondary component is zero.

It is already shown that, most probably, magnetic reconnection is responsible for the origin of secondary Gaussian. Further, the unresolved motions (i.e. non-thermal velocities) of the secondary Gaussian are slightly higher than the non-thermal motions of the primary Gaussian. Hence, the secondary physical process (i.e. magnetic reconnection) plays an essential role in the dynamics of TR in QS and CH as it has higher non-thermal motions (i.e. the dominating physical process in the TR). Finally, we can say that this dominating secondary process may hold the key to the heating and mass cycle of the solar TR in QS and CH.

Unlike QS and CH, on average the secondary component is redshifted. Hence, in the case of BPs, the magnetic reconnection for secondary Gaussian is happening above the formation height of the Si iv line. Hence, the height of the magnetic reconnection (responsible for the secondary component) in BPs is likely to be different from QS and CHs. Then, we can say that primary as well as secondary Gaussian are redshifted in BPs. Further, we mention that the non-thermal mentions are much stronger in the BPs than in QS and CH.

## ACKNOWLEDGEMENTS

We gratefully acknowledge the anonymous reviewer for his constructive comments that improved the manuscript. We appreciate the utilization of data from Interface Region Imaging Spectrograph (IRIS), Atmospheric Imaging Assembly (AIA), and Helioseismic and Magnetic Imager (HMI). IRIS is a NASA small explorer assignment developed and functioned by LMSAL with assignment operations performed at NASA Ames Research Center and major portions to downlink connections funded by ESA and the Norwegian Space Center. AIA and HMI are appliances on board Solar Dynamics Observatory (SDO), an assignment for NASA's Living With a Star plan.

## DATA AVAILABILITY

The three data sets for IRIS and AIA are available at: https://iris.lmsal.com/search/. Also, the cotemporal-spatial HMI magnetograms can be downloaded from JSOC: http://jsoc.stanford.edu/ajax/exportdata.html

## APPENDIX A: DATA SET 2 – 2014 JULY 24

In this appendix, we have displayed the same figures corresponding to the second data set.

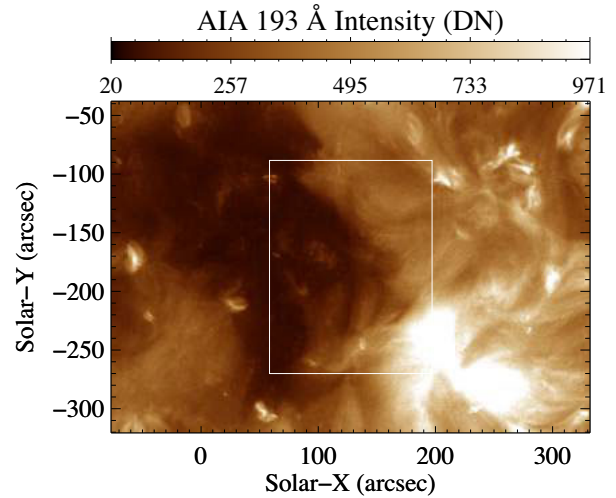

**Figure A1.** Same as Fig. 1 but for data set 2.

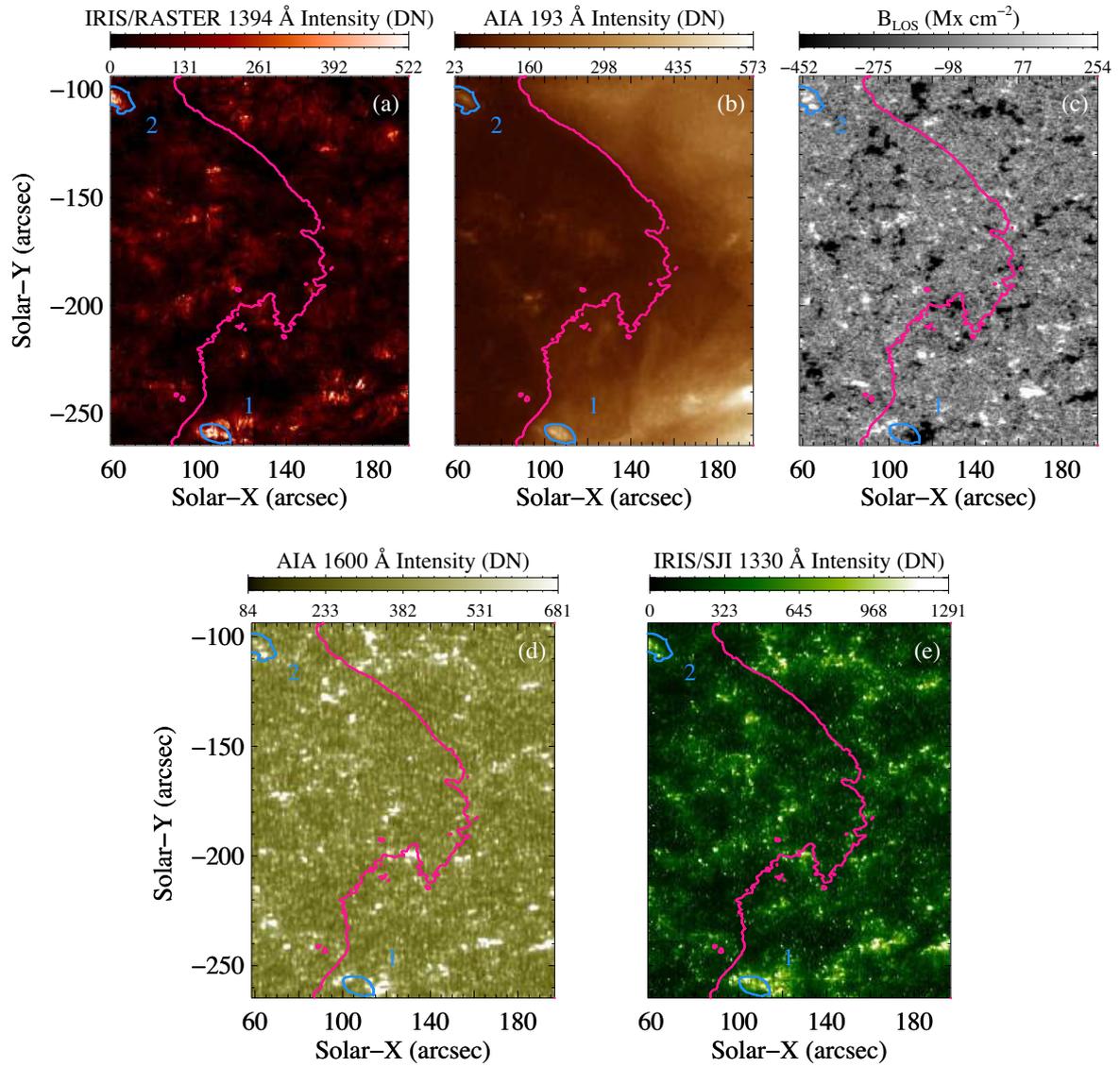

**Figure A2.** Same as Fig. 2 but for data set 2.

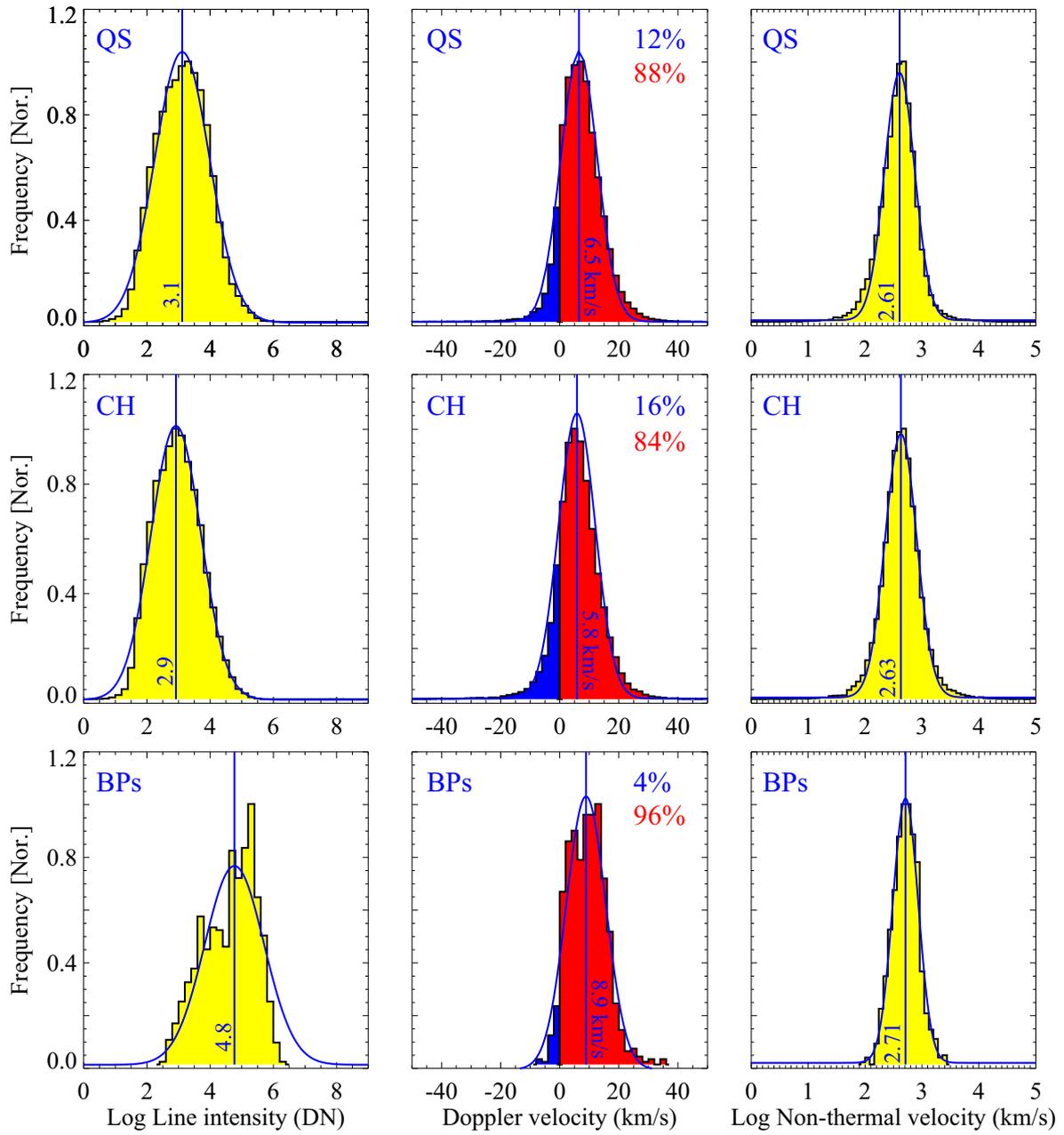

**Figure A3.** Same as Fig. 4 but for data set 2.

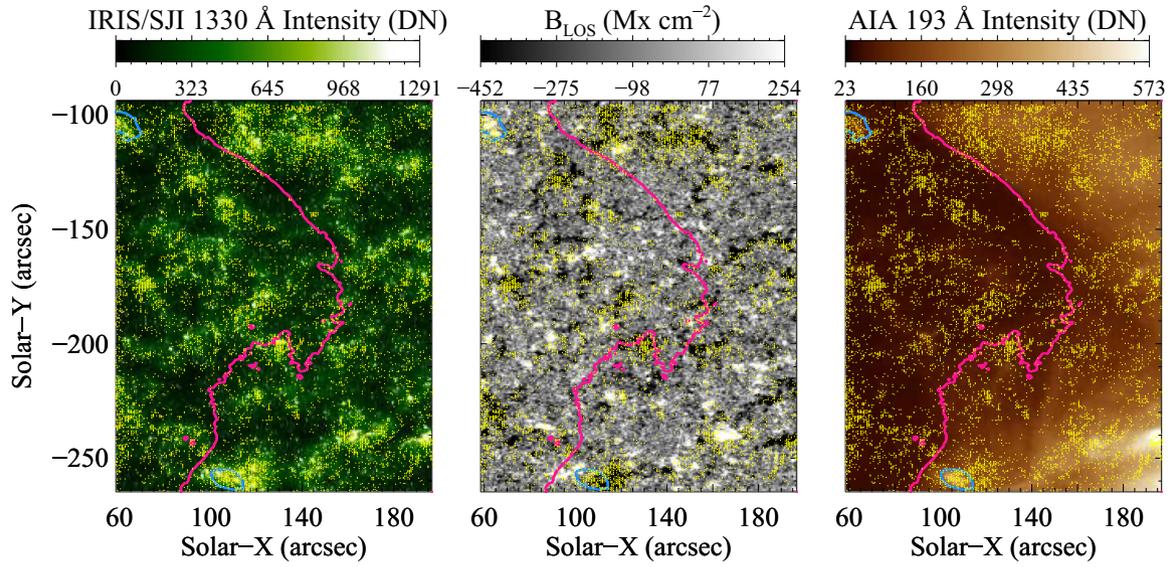

**Figure A4.** Same as Fig. 5 but for data set 2.

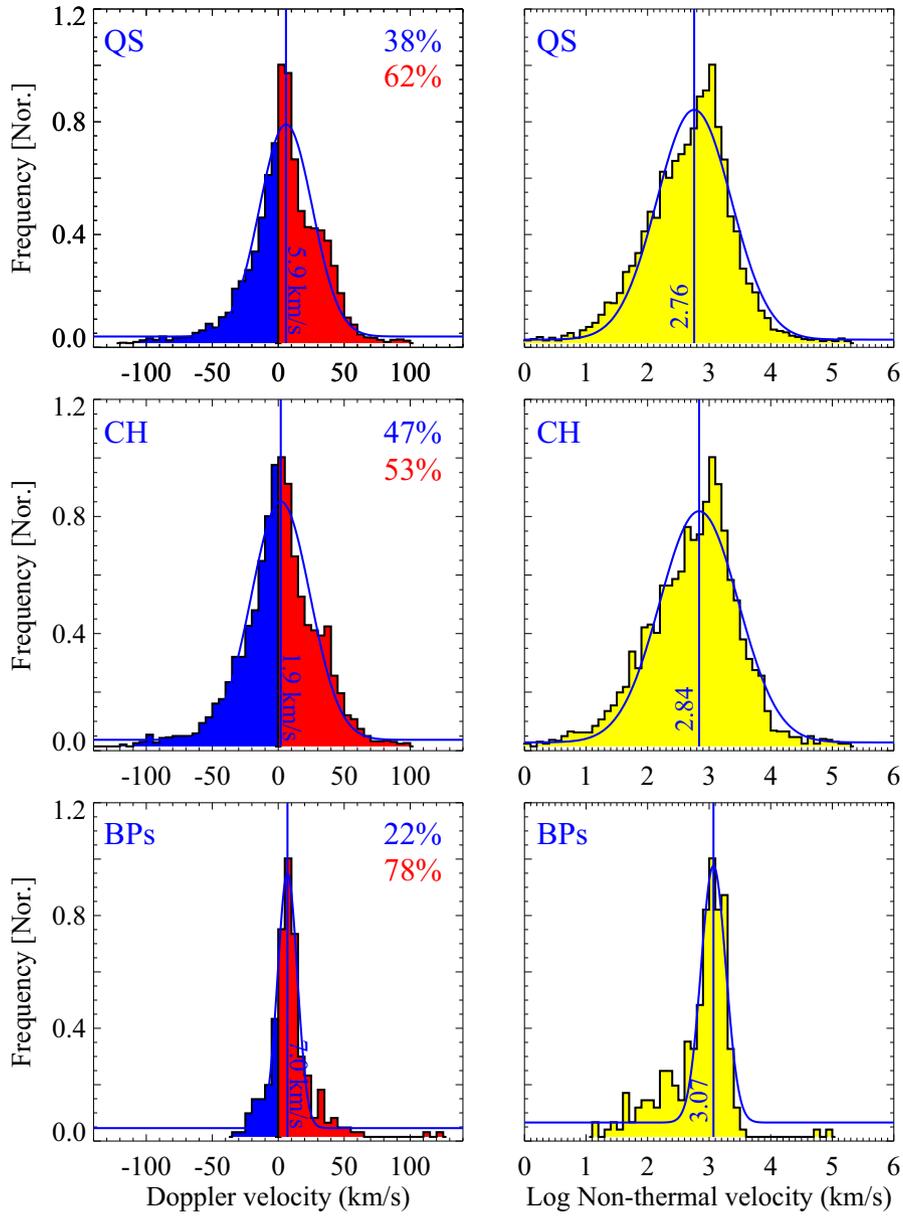

**Figure A5.** Same as Fig. 6 but for data set 2.

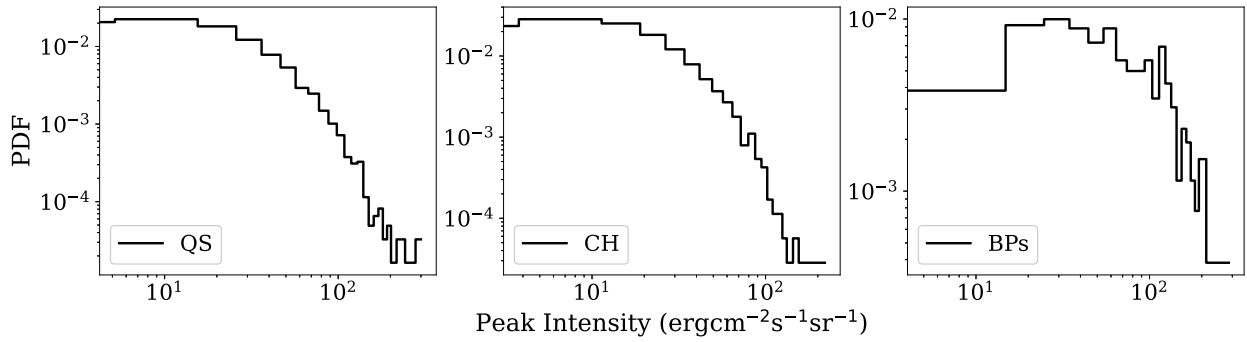

**Figure A6.** Same as Fig. 7 but for data set 2.

# APPENDIX B: DATA SET 3 – 2014 JULY 26

In this appendix, we have displayed the same figures corresponding to the third data set.

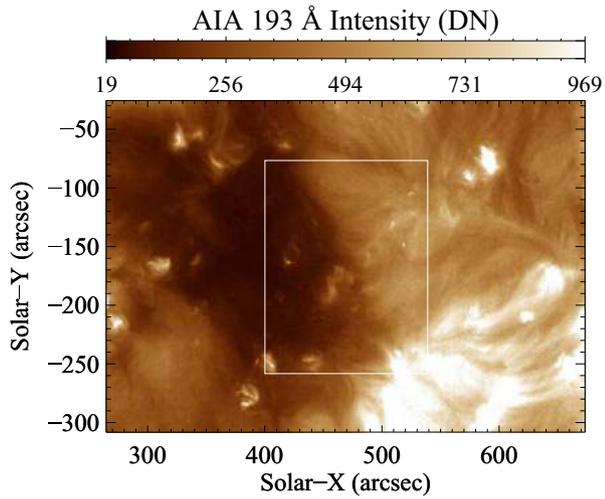

**Figure B1.** Same as Fig. 1 but for data set 3.

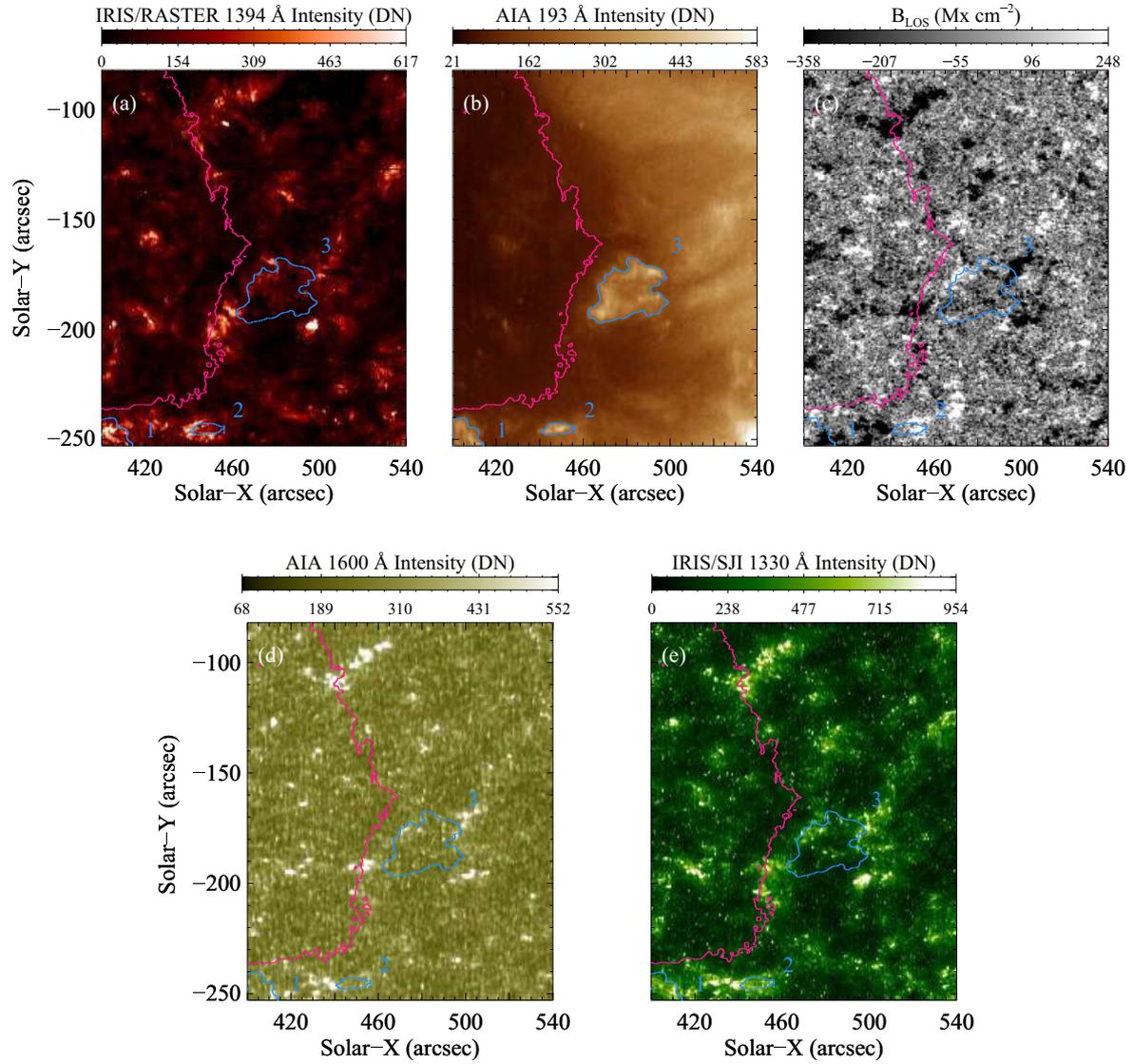

**Figure B2.** Same as Fig. 2 but for data set 3.

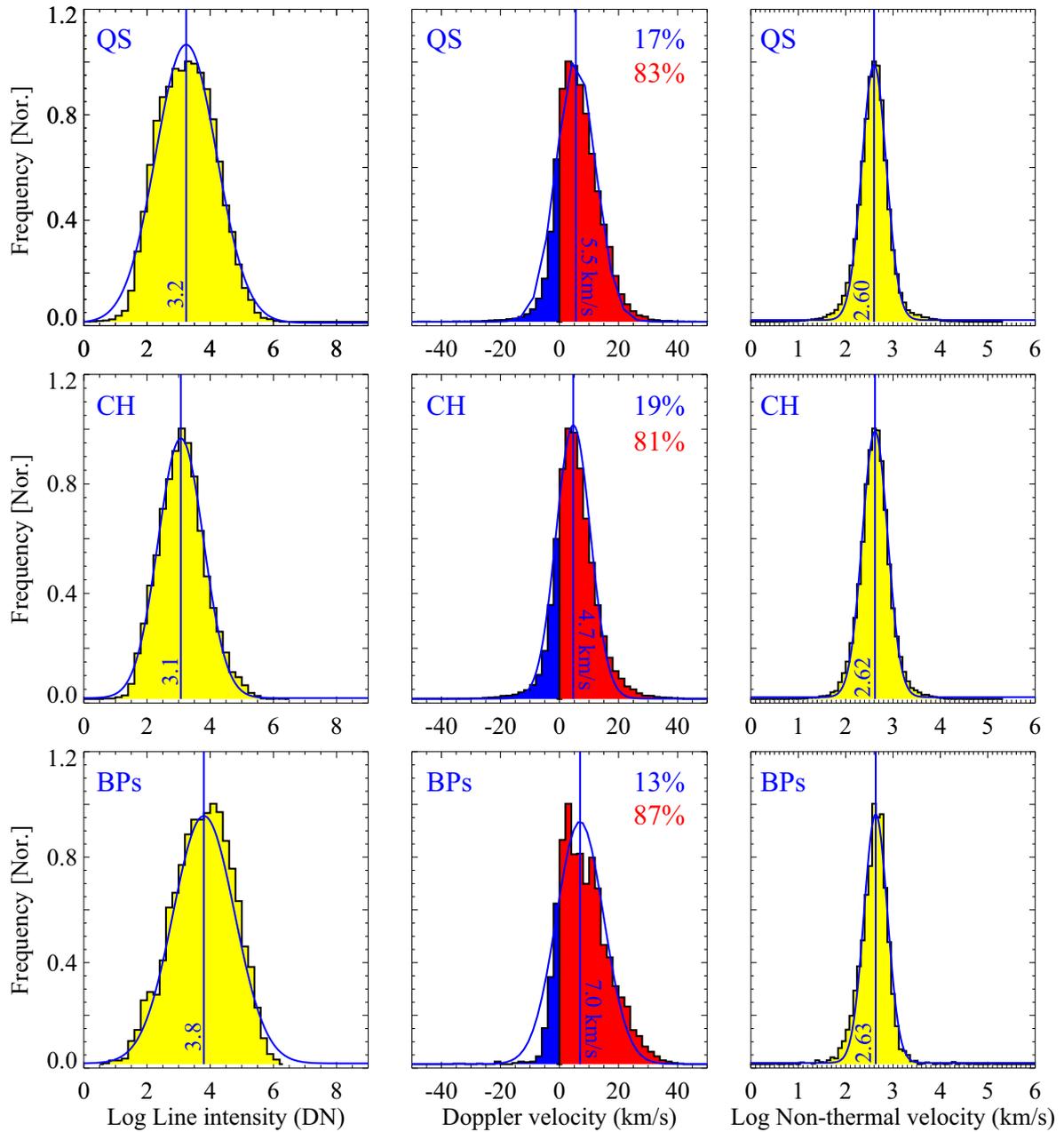

**Figure B3.** Same as Fig. 4 but for data set 3.

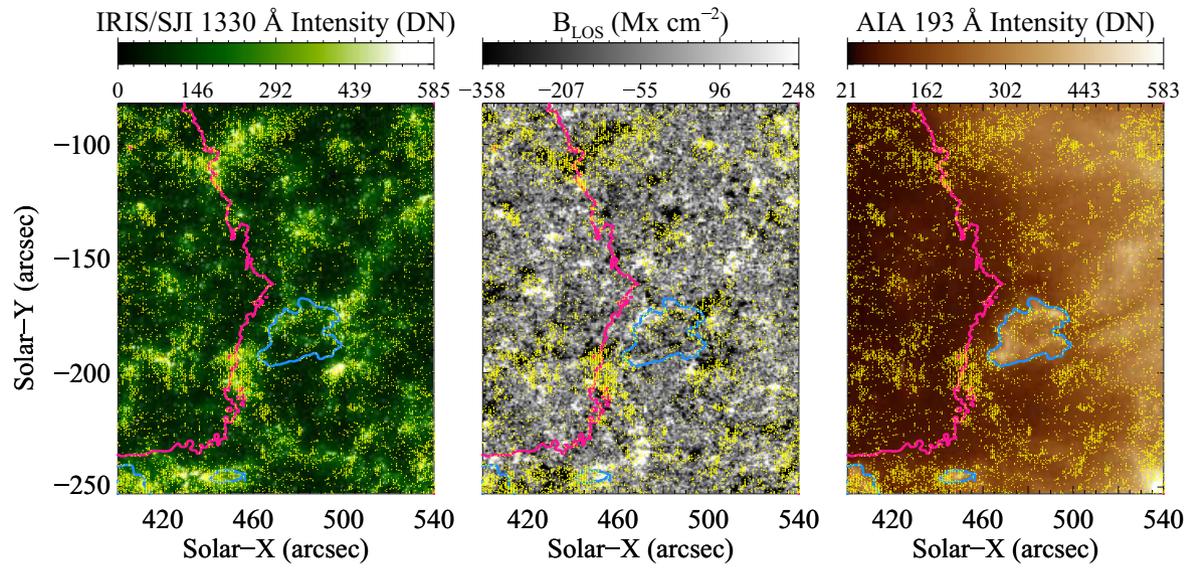

**Figure B4.** Same as Fig. 5 but for data set 3.

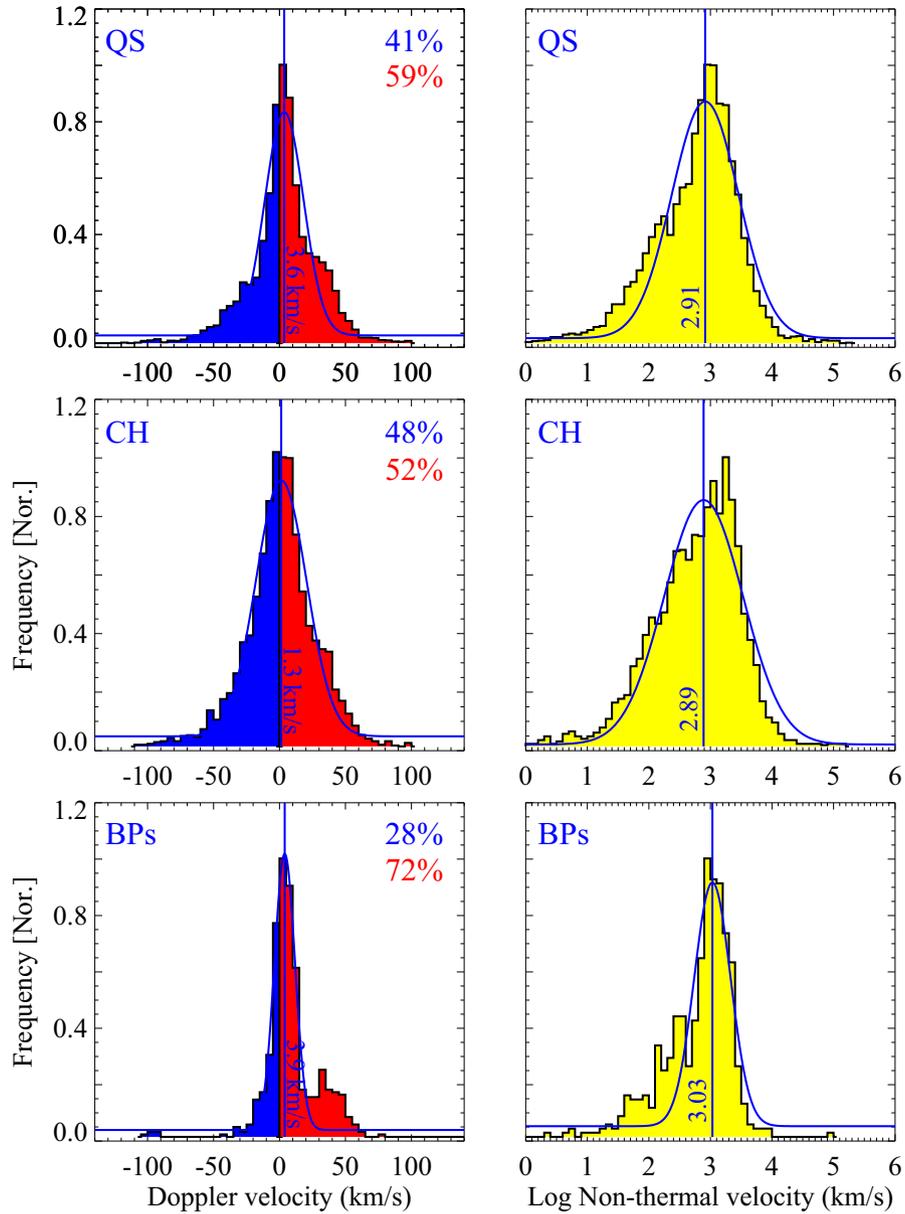

**Figure B5.** Same as Fig. 6 but for data set 3.

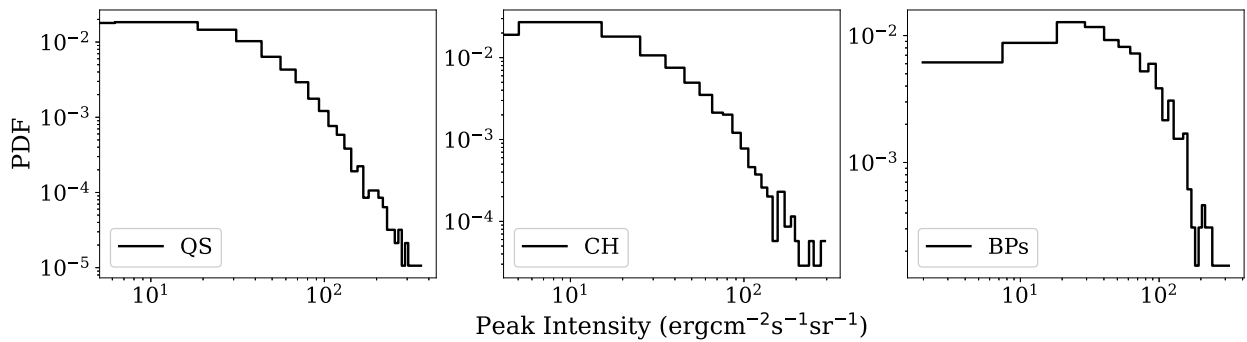

**Figure B6.** Same as Fig. 7 but for data set 3.

This paper has been typeset from a TeX/LaTeX file prepared by the author.